\documentclass[pageno]{jpaper}

\newif\ifisdraft
\isdraftfalse

\usepackage[normalem]{ulem}

\usepackage[T1]{fontenc}
\usepackage[utf8]{inputenc}

\usepackage{amsthm}
\usepackage{scalerel, amssymb}
\usepackage{amsmath}

\usepackage{listings}

\usepackage{hyperref}
\usepackage{xspace}

\usepackage{color}
\definecolor{bluekeywords}{rgb}{0.13,0.13,1}
\definecolor{greencomments}{rgb}{0,0.5,0}
\definecolor{redstrings}{rgb}{0.9,0,0}

\usepackage{tikz}
\usetikzlibrary{calc}
\usetikzlibrary{automata,positioning}
\usepackage[all,cmtip]{xy}

\ifisdraft
\newcommand{\redcomment}[1]{\textcolor{red}{\textbf{#1}} }

\else
\newcommand{\redcomment}[1]{ }

\fi

\newcommand{\RA}[1]{\redcomment{RA: #1}} 
\newcommand{\LH}[1]{\redcomment{LH: #1}}

\newcommand{\etal}{\emph{et al.}\xspace}

\newcommand{\ei}{{\it i)}\xspace}
\newcommand{\eii}{{\it ii)}\xspace}
\newcommand{\eiii}{{\it iii)}\xspace}

\makeatletter
\newcommand{\customlabel}[2]{%
  \protected@write \@auxout {}{\string \newlabel {#1}{{#2}{\thepage}{#2}{#1}{}} 
  }%
  \hypertarget{#1}{#2}
}
\makeatother

\newcounter{invariant}
\newcommand{\invariant}[3] {
  \refstepcounter{invariant}
  \begin{center}
    \begin{tabular}{|p{0.95\columnwidth}|}
      \hline
      \textbf{\customlabel{inv:#1}{Invariant I\theinvariant}\ (#2)}
      #3
      \\\hline
    \end{tabular} 
  \end{center}
}

\newcounter{assumption}

\newcounter{rights}
\newenvironment{rightsgroup}
  {\begin{center} \begin{tabular}{|p{0.95\columnwidth}|}\hline}
  {\hline\end{tabular} \end{center}}
\newcommand{\rightsgroupentry}[3] {
    \vspace{-0.6em}
   \refstepcounter{rights}\textbf{\customlabel{right:#1}{Right R\therights}\ 
   (#2)} #3\\}

\lstdefinestyle{code}
{language=C,
	basicstyle=\tiny,
	showspaces=false,
	showtabs=false,
	breaklines=true,
	showstringspaces=false,
	breakatwhitespace=true,
	escapeinside={(*@}{@*)},
	commentstyle=\color{greencomments},
	keywordstyle=\color{bluekeywords}\bfseries,
	stringstyle=\color{redstrings},
	basicstyle=\ttfamily,
	morekeywords={uint8_t, uint16_t, uint32_t, uint64_t, size_t, err_t, 
	regionid_t},}

\ifisdraft

\usepackage[framemethod=default]{mdframed}

\mdfdefinestyle{inv}{%
	linecolor=black,linewidth=2pt,%
	topline=false,
	leftline=false,
	innertopmargin=1pt,
	skipbelow=\topskip, 
	skipabove=\baselineskip,
}

\mdfdefinestyle{defi}{%
	linecolor=black,linewidth=2pt,%
	frametitlerule=true,%
	frametitlebackgroundcolor=gray!20,
	innertopmargin=3pt,
	skipbelow=\topskip, 
	skipabove=\baselineskip,
	nobreak=true
}

\mdtheorem[style=defi]{defi}{Definition}
\mdtheorem[style=defi]{op}{Operation}
\mdtheorem[style=inv]{inv}{Invariant}

\fi

\begin{document}
\newcommand{\system}{\textit{Barrelfish/MAS}\xspace}

\title{A Least-Privilege Memory Protection Model for Modern Hardware}

\author{Reto Achermann, Nora Hossle, Lukas Humbel, Daniel Schwyn, David Cock, Timothy Roscoe\\
Systems Group, Department of Computer Science, ETH Zurich}





\date{}
\maketitle

\begin{abstract}

We present a new least-privilege-based model of addressing on which to
base memory management functionality in an OS for modern computers
like phones or server-based accelerators.  Existing software
assumptions do not account for heterogeneous cores with different
views of the address space, leading to the related problems of
numerous security bugs in memory management code (for example
programming IOMMUs), and an inability of mainstream OSes to securely
manage the complete set of hardware resources on, say, a phone
System-on-Chip.

Our new work is based on a recent formal model of address translation
hardware which views the machine as a configurable network of address
spaces.  We refine this to capture existing address translation
hardware from modern SoCs and accelerators at a sufficiently fine
granularity to model minimal rights both to access memory and
configure translation hardware.  We then build an executable
specification in Haskell, which expresses the model and metadata
structures in terms of partitioned capabilities.  Finally, we show a
fully functional implementation of the model in C created by extending
the capability system of the Barrelfish research OS.

Our evaluation shows that our unoptimized implementation has
comparable (and in some cases) better performance than the Linux
virtual memory system, despite both capturing all the functionality of
modern hardware addressing and enabling least-privilege, decentralized
authority to access physical memory and devices.

\end{abstract}

\section{Introduction}
\label{sec:introduction}

Both modern, fully-verified operating systems and traditional
production-quality kernels rely on a model of memory addressing and
protection so simple it is rarely remarked on: RAM and devices reside
at unique addresses in a single, shared physical address space, and
all cores have homogeneous memory management units which translate
from a virtual address space into these physical addresses.  These
MMUs are all configured by a single monolithic kernel.

Unfortunately, this model bears little relation to modern hardware.
Modern platforms like phone SoCs violate the assumption of a single
physical address space.  A modern computer is, in reality, a network of
address spaces with ad-hoc address translation functions between them,
many configurable by sufficiently-privileged system software.  Access
to memory is performed by a variety of heterogeneous cores and I/O
devices from different points in this network.  Simply configuring a
given platform correctly to maintain the assumption of a single
physical address space on which the verification is based on is, by itself,
a complex and error-prone process.

The result is that traditional kernels suffer from numerous (and
continuing) security bugs arising from incorrect assumptions about memory
addressing in the system, while correctness proofs for verified
kernels are cast into doubt by the existence of ``cross-SoC''
attacks.

Moreover, centralized authority over all memory access does not
accommodate features like the secure co-processors and management engines
standard in modern PCs as well as phone platforms.  Authority to grant
access to memory and devices needs to be decentralized, and this
decentralization represented in the specification of the OS itself.

This paper develops an alternative model of hardware memory
addressing, protection, and authorization
which captures the richness, complexity, and diversity of modern
hardware platforms.  Our model can serve as a basis for formal
verification of system software but also as an informal basis for
designing correct memory management functionality.   

The model is based on two guiding principles: \emph{completeness},
meaning that we capture the full semantics of real addressing hardware
without simplifying assumptions, and \emph{least-privilege}, meaning
that we represent individual authority to both access memory and modify
translations at as fine a granularity as allowed by the hardware.

In the next section we elaborate on the mismatch between modern
hardware and OS designs, and existing efforts to address it.  In
Section~\ref{sec:methodology} we review the recent related work on
which this paper builds, and lay out our methodology. 

Our first contribution, in Section~\ref{sec:design}, is
development of the model itself.  We start from an abstract
model of memory addressing and progressively refine it
until it captures the salient features of modern memory hardware,
including the rights to modify translations in multi-level page tables
and custom protection units.  We build an executable spec in
Haskell which serves as a basis for an implementation in a real OS.

In Section~\ref{sec:implementation} we describe how Linux might be
extended with a subset of our model (foregoing least-privilege and
centralizing authority in the kernel), and present a full
implementation of the model by extending the capability system in the
Barrelfish research OS.  Our implementation runs on real hardware and
can manage protection rights on a variety of hardware platforms.  We
also discuss the minimal overhead it incurs for metadata and bookkeeping.

In Section~\ref{sec:evaluation} we evaluate the performance of this
memory system and show that, despite the richer and more faithful
view of hardware it embodies, it provides
comparable performance to the highly optimized, but less
functional, Linux virtual memory system on identical hardware.


This paper also has a set of \emph{non-goals}.  Firstly, we do not
present any formally-verified OS software; our goal is rather to show
a model which can be used as a replacement for the over-simplified
addressing models currently used in the proofs for verified systems
like seL4 and CertiKOS.

Second, we develop no new memory subsystem for
Unix-like OSes.  As we note below, reasoning
about the correctness and security of a modern computer requires going
beyond a Linux kernel to capture co-processors and intelligent devices.
We sketch in section~\ref{sec:impl:linux} how a simplified
version of our model might be retrofitted to Linux.

Finally, this is not a bug-finding paper.  We do not aim to find
problems in existing OS code (though we cite
numerous examples from other work).  Instead, we lay the foundations
for a more faithful view of the hardware on which to base better
system software. 

\section{Motivation and Related Work}
\label{sec:background}

We first review the implicit model of memory addressing used
by existing OSes, and then explain with concrete examples why it no
longer reflects hardware reality.  We discuss the implications of this
for both new, formally verified OSes and traditional kernels like
Linux, and the limitations of existing approaches to the
problem in both kinds of OS.  

\subsection{The traditional view of memory}

Address translation is a fundamental technique in computing,
enabling relocation, demand paging, machine virtualization 
(either via processes or full virtual machines), shared memory,
inter-process protection, and much other functionality.

Typically, the key abstraction employed is a \emph{virtual address
  space}, accesses to which are translated into addresses within a unique,
machine-wide \emph{physical address space} by hardware mechanisms
(TLBs, multi-level page tables, etc.).

Physical memory addresses can thus be used as unambiguous, system-wide
identifiers for memory and devices, and so are also used to keep track
of access rights: Linux maintains such a data-structure for each page
or frame of physical memory, while some microkernels like
seL4~\cite{Elkaduwe:2008:VPM, Klein:2009:SFV} and
Barrelfish~\cite{Baumann:2009:MNO, Gerber:2018:APA} use a capability
system~\cite{Levy:1984:CCS:538134} to represent physical memory regions with
access rights.

An OS must configure translation hardware and maintain these data
structures to ensure correct and secure operation. For example, user
programs should only be able to load and store to physical resources
(memory, or memory-mapped I/O devices) the OS has granted them access
rights to.

\subsection{Hardware doesn't conform to this view}

Unfortunately, modern hardware platforms violate the assumptions in
the traditional model above.  They are composed of multiple, heterogeneous cores 
and devices each of which can issue accesses to byte-addressable 
memory resources such as DRAM, non-volatile memory or device
registers.  Worse, there is no single ``reference'' physical address
space~\cite{Gerber:2015:YPP}.  Instead, a network of address spaces or
buses is connected by address translation units which ``routes''
accesses through the network.

This breaks most of the assumptions of the classical model: different
cores and devices translate their virtual addresses into different
physical address spaces, physical addresses can no longer be used as
global identifiers without further scoping,  address aliasing is not only
possible but likely, and finally, software with access to translation
units can reconfigure the \emph{physical} address space underneath the
systems' MMUs.

For example, the Xeon Phi co-processor~\cite{Intel:2014:XeonPhi}
implements a ``system memory page table'' which further translates
physical (post-MMU) addresses from the accelerator cores into the
host's PCI address space using a single, shared register array where
each register controls the translation of a fixed 16GB page in the
Xeon Phi's ``physical'' address space.

Such additional layers of translation are commonplace in phone
Systems-on-Chip like the NXP iMX8~\cite{nxp:2019:imx8xrm}, Texas
Instruments OMAP~\cite{ti:2011:omaptrm}, and NVIDIA
Parker~\cite{nvidia:2017:parker} processors.  Such SoCs contain a
variety of different processors with different physical address
spaces, which overlap and intersect~\cite{Gerber:2015:YPP}.  This is a
deliberate, rational design choice -- for example, it is important
that a secure co-processor holding encryption keys has private memory
that cannot be accessed from application cores, even in kernel mode.

I/O memory management units (IOMMUs, or System MMUs) translate
addresses generated by accelerators and DMA-capable devices into a
``canonical'' system-wide physical address space.  This allows
user-space programs to share a virtual address space with a context on
the device, but impose a further complexity burden on the underlying
OS which must now ensure that IOMMUs are always correctly
programmed.  This code is fraught with complexity and consequent bugs
and vulnerabilities, as it is also intended to provide protection from
malicious memory accesses~\cite{Morgan:2016:BIP, Morgan:2018:IPIO,
  Markuze:2016:TIP, Markettos:2019:TEV}.
The problem is likely going to get worse with the proliferation of IOMMU
designs built into GPUs, co-processors, and intelligent NICs. 

OpenCL's Shared Virtual Memory extends the global memory region into
the host memory region using three different
types~\cite{Khronos:2015:OpenCL}.  Similarly, nVidia's
CUDA~\cite{nVidia:2013:UMCUDA} or HSA~\cite{HSA:2016:RPR} provide a
unified view of memory.
The same concerns apply here: the
complexity of maintaining a shared virtual address space is pushed to
system software, but remains.

Even memory controllers can violate the traditional model.
Hillenbrand \etal~\cite{Hillenbrand:2017:MPM} reconfigure memory
controller configurations from system software to provide DRAM aliases
for mitigating the performance effects of channel and bank
interleaving.  Proposals for ``in-memory'' or ``near-data''
processing~\cite{Patterson:1997:CIR} raise further questions for OS
abstractions~\cite{Barbalace:2017:TTO} and require a way to
unambiguously refer to memory regardless of which module accesses it.

\subsection{Implications for current OS designs}

Correctness arguments about OS code therefore rely on assumptions
about the hardware that no longer hold.  Proofs for the seL4
microkernel~\cite{Klein:2009:SFV} assume a single, fixed, physical
address space without other translation hardware, and provide no
guarantees of safety in the presence of other cores or incorrectly
programmed DMA devices.  CertiKOS~\cite{Gu:2016:CEA} proves functional
correctness based on a model of memory accesses to abstract regions of
private, shared or atomic memory, but again provides no proof in the
presence of other translation units and heterogeneous cores.  Even
work on verifying memory consistency in the presence of translation
only considers the simple case of virtual-to-physical
mappings~\cite{Romanescu:2010:SDV}.

Proofs aside, the difficulty of getting complex memory addressing
right in an OS is shown by the steady stream of related bugs and
vulnerabilities in Linux~\cite{Huang:2016:ESL}, for example ignoring
holes in huge pages (CVE-2017-16994), miscalculation of the number of
affected pages (CVE-2014-3601), access rights for data pages
(CVE-2014-9888), interactions of virtually mapped stack with DMA
scatter lists (CVE-2017-8061), handling of shadow page tables
(CVE-2016-3960). Moreover, miscalculations, misinterpretations or
underflows of addresses and offsets, (Linux commits 9d8c3af3160, 7655739143,
29a90b708 and 5016bdb79), mixing up memory addresses with MSI-X interrupt
ranges (Linux: 17f5b569e09cf) and IOMMU address space allocations
(Linux: a15a519ed6e) cause unexpected behavior, crashes or memory
corruption.

Faced with the complexity of hardware, a number of ad-hoc point
solutions have appeared for specific cases, primarily GPUs, such as
VAST~\cite{Lee:2014:VIL} which uses compiler support to dynamically
copy memory to and from the GPU and
Mosaic~\cite{Ausavarungnirun:2017:MGM}, which provides support for
multiple sizes of page translation in a shared virtual address space
between CPU and GPU.
In DVMT~\cite{Alam:2017:DVM}, applications request physical frames from the OS 
that have specified properties. The system allows applications to customize how 
the virtual-to-physical mapping is set up by registering a TLB miss handler for 
the special DVMT range.
The CBuf\cite{Ren:2016:CES} system globally manages virtual and physical memory 
focusing on efficient sharing and moving data between protection domains. CBuf 
unifies shared-memory, memory allocation and system-wide physical memory 
allocation.

All these approaches aim to simplify user code, at the cost of OS
complexity.  In contrast, our work is a response to this complexity:
the central OS abstraction of a single, shared, global physical
address space, combined with straightforward translations to it from
virtual address spaces, is inadequate for a secure and reliable OS
running on modern hardware.  We need a richer model of addressing, and
this paper is based on one which views address spaces as nodes in a
network of translation units.

\section{Methodology}
\label{sec:methodology}

Our new model builds on the existing \emph{decoding
net} model of Achermann~\etal~\cite{Achermann:2017:FMA,Achermann:2018:PAR},
which has been shown to provide a precise formal model of many of the sorts of
systems we consider in this work: Multi-socket NUMA systems, ARM SoCs, plug-in
accelerators, etc.

Achermann~\etal~model the addressing structure of a system as
a directed graph, where nodes represent (virtual or physical) address spaces
or devices (including RAM), and edges the translation of
\emph{AS-local} addresses into other ASs or devices.  The graph is a set of
nodes, defined as an abstract datatype so:
\begin{align*}
\textit{name} &= \textrm{Name}\ \textit{nodeid}\ \textit{address}\\
\textit{node} &= \textrm{Node}\ \textbf{accept}::\{\textit{address}\}\  \\
               &\phantom{{}=\textrm{Node}\ }\textbf{translate} ::
                                \textit{address}\rightarrow \{name\}
\end{align*}
Their model distinguishes \emph{local} names ($\textit{address}$), relative to
some address space, and \emph{global} names ($\textit{name}$), which qualify a
local name with its enclosing address space.  Each node may \textbf{accept}
a set of (local) addresses, and/or \textbf{translate} them to one or more
global names (addresses in other address spaces).

This existing model is a long way from being a basis for an
operational system.  In Section~\ref{sec:model} we add two important
features: dynamic configuration of the \textbf{translate} function
which captures how real translation units can be programmed, and
\emph{rights} corresponding to the ability for software processes to
configure such units.
We model the complex network of interacting address spaces, 
identify and label the necessary divisions of authority as finely as
possible, following the principle of least privilege.

We adopt a methodology strongly influenced by the successful
combination of \emph{refinement} and \emph{executable specification}
used in the seL4 project.

Specifically, we begin by identifying all relevant \emph{objects} (page
tables, address spaces, \ldots), the \emph{subjects} that manipulate them
(processes, the kernel, devices, \ldots), and which \emph{authority} each
subject exercises over an object (e.g.~in mapping a frame to a virtual
address).  These are expressed in an \emph{access-control matrix} (following
Lampson~\cite{Lampson:1974:Protection}) which forms our \emph{abstract
specification}, analogous to the high-level \emph{security policy} (integrity)
shown to be refined (correctly implemented) all the way down to compiled
binaries for seL4~\cite{Sewell:2011:Integrity}.

Again, as in seL4~\cite{Cock:2008:SMS}, we next develop an executable
specification in Haskell (see Section \ref{sec:haskell}), expressing
subjects, objects, and authority as first-class objects, permitting rapid
prototyping without giving up strong formal semantics.  Correspondence between
abstract and executable models is thus far by inspection and careful
construction.

Finally, we show (again with precedent~\cite{Winwood:2009:MG}) that the
executable model (and hence the abstract model) permits multiple
high-performance implementations: In the Barrelfish OS (expressing
\emph{rows} of the access matrix with capabilities, see Section
\ref{sec:impl:barrelfish}), and in Linux (collapsing distinct authorities held
by the kernel, and taking \emph{columns} as access-control lists, see Section
\ref{sec:impl:linux}).  Barrelfish and seL4 have closely related capability-based
resource management and authorization systems and our implementation
transfers naturally to seL4; Barrelfish is \emph{currently} a better
platform for our work, due to its support for multiprocessing
and heterogeneous hardware.

By adopting a proven methodology, we can be confident that the resulting
artifact is compatible with an seL4-style verification, and could thus serve
as a more accurate replacement for the hardware model underlying the seL4 or
CertiKOS proofs.  Simultaneously, by careful selection of an abstract model
(the access-control matrix) and through the use of refinement, our model is not
specific to a particular implementation. 

\section{Model}
\label{sec:design}


We derive our abstract, formal model from the existing decoding net model in
two steps.  First, we extend the model to include dynamic behavior (updating
translations), and express the required authority using an access-control
matrix.  Second, we build a (still relatively abstract) executable
specification in Haskell, allowing us to reason concretely about
implementation trade-offs.

\subsection{Authority and Dynamic Behaviour}
\label{sec:model}

Decoding nets are static: they represent the \emph{current} state of the
system.  To describe the dynamic behavior of a system, we add an abstraction
above the decoding net, consisting of a set of (dynamic) \emph{address
spaces}.  The state of the system is then expressed as a function from address
space, to the mapping node representing its current configuration:
\begin{displaymath}
\textit{configuration} = \textit{address space} \rightarrow \textit{node}
\end{displaymath}
We can then express the \emph{configuration space} of an address space, as a
set of possible configurations:
\begin{displaymath}
\textit{config space} = \textit{address space} \rightarrow \{\textit{node}\}
\end{displaymath}
The configuration space of a page table in a system with 4kiB translation
granularity would, for example, only include nodes that map all addresses in
any naturally-aligned 4kiB region contiguously.
We will use the configuration space to express allowable system
states according to a security property.

At this level of abstraction, state transitions are simply changes in the
current configuration of the address spaces:
\begin{align*}
\texttt{ModifyMap}\ &:: \textit{address} \rightarrow
                        \textit{name} \rightarrow \\
                    &\phantom{{}::} \textit{configuration} \rightarrow
                                    \textit{configuration}
\end{align*}
%
%

%
%
%
%
%


\subsubsection*{Authority}

\begin{figure}
  \includegraphics[width=\columnwidth]{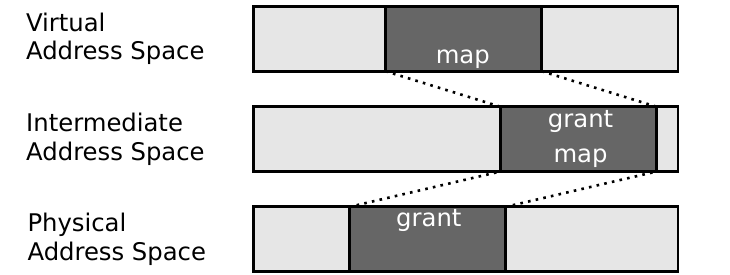}
  \vspace{-0.5cm}
  \caption{Mappings between 
  address spaces showing grant and map rights of mapped segments.}
  \label{fig:model:single-address-space}
\end{figure}

Consider~\autoref{fig:model:single-address-space}, representing the general
case of an update to an intermediate address space (for example the
intermediate physical address, IPA, in a two-stage translation system).  We
identify two distinct rights (authorities): The \textbf{map} right, or the
right to change the meaning of an IPA by changing its mapping; and the
\textbf{grant} right, or the right to grant access (by mapping) to some range
of physical addresses.

\begin{figure}
  \includegraphics[width=\columnwidth]{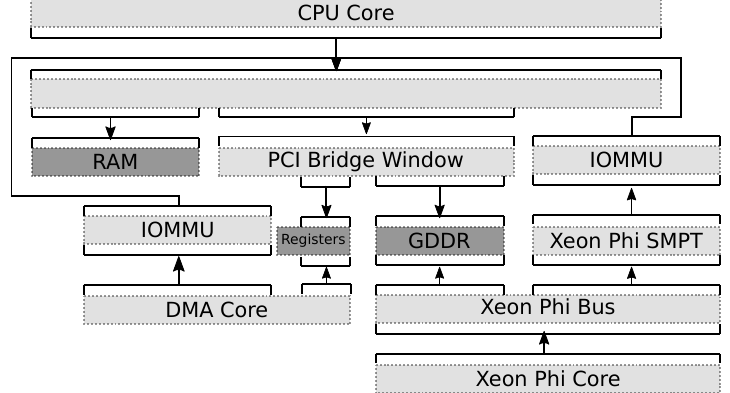}
  \vspace{-0.5cm}
  \caption{Address spaces in a system with two PCI devices}
  \label{fig:model:multiple-address-spaces}
\end{figure}

These two rights do not necessarily go together.

Consider~\autoref{fig:model:multiple-address-spaces}, showing the
address-space structure of a system with two PCI devices: a DMA engine and an
Intel Xeon Phi co-processor.  Imagine that we wish to establish a shared
mapping to allow a process on a Xeon Phi core to receive DMA transfers
(e.g.~network packets) into a buffer allocated to it in the on-board GDDR.

The process `owns' the buffer, and has the ability to call \texttt{recv()},
triggering a DMA transfer.  We interpret this as the process having the right
to \textbf{grant} access (temporarily) to the DMA core.  The user-level
process, however, clearly should not have the ability to modify the IOMMU
mappings of the DMA core at will (or its own, for that matter).  That is, it
does not have the \textbf{map} right on the relevant address space.

What is needed is some agent (hereafter a \emph{subject}, in standard
authority-control terminology) with both the \textbf{grant} right on the
buffer \emph{object}, and the \textbf{map} right on the address space
\emph{object}.  In a traditional monolithic kernel, both these rights are held
(implicitly) by the kernel, which exercises them on behalf of the subjects.
It is up to the kernel to maintain accurate bookkeeping to determine whether
any such request is safe, typically using an ACL (access-control list)
i.e.~authority tied to the \emph{subject}.

In a microkernel such as seL4 or Barrelfish, these rights are represented by
capabilities, handed explicitly to one \emph{subject}, in order to authorize
the operation.  In this case, authority is tied to the \emph{object}.  These
are equivalent from the perspective of access control, differing only on
implementation detail: In both cases, the same two basic authorities are
present.

\begin{rightsgroup}
  \rightsgroupentry{grant}{Grant}{\\
  The right to insert \emph{this} object into \emph{some} address space}
  \rightsgroupentry{map}{Map}{\\
  The right to insert \emph{some} object into \emph{this} address space}
\end{rightsgroup}

Note that the 'virtual' and 'physical' address spaces 
of~\autoref{fig:model:single-address-space} can be viewed as special cases of
an intermediate address space: A top-level 'virtual' address space is simply
one to which \emph{nobody} has a \textbf{grant} right, and a 'physical'
address e.g.~DRAM is one to which there exists no \textbf{map} right.

\begin{table}
  \begin{footnotesize}
  \begin{center}
    \begin{tabular}{l|ccc}
      \emph{subject}/\emph{object}  & DMA IOMMU    & buffer \\
      \hline \\
      IOMMU driver                  & \textbf{map} &        \\
      Xeon Phi process              &              & \textbf{grant}
    \end{tabular}
  \end{center}
\end{footnotesize}
  \vspace{-1em}
  \caption{Access control matrix of the Xeon Phi example}
  \label{tab:model:acm}
\end{table}

The standard representation of authority in systems is an access
control matrix~\cite{Lampson:1974:Protection}, such as that
of~\autoref{tab:model:acm}.  This can be read in rows: The IOMMU
driver has the \textbf{map} \emph{capability} to the IOMMU address
space, and the process the \textbf{grant} capability to the buffer.
Alternatively, reading down the columns gives the ACLs: the IOMMU
records \textbf{map} \emph{permission} for the driver, and for the
buffer is recorded a \textbf{grant} \emph{permission} for the process.

This access control matrix on maps and grants is our abstract model.  A system
is correct (secure) \emph{statically}, if its current configuration is
consistent with the access control matrix.  It is secure \emph{dynamically} if
any possible transition, beginning in a secure state, must leave the system in
a secure state.

%





\subsection{Executable Specification}
\label{sec:haskell}

Thus far we have expanded upon the existing decoding net model, giving us a
dynamic access-control matrix formulation of the system's correctness
property.  Next, we implement a (still abstract) \emph{reference
monitor}~\cite{Anderson:1972:reference_monitor} in Haskell, to aid in rapid
prototyping of both model and implementation, and as an intermediate step in
the process of \emph{refinement} from abstract specification to operational,
high-performance implementation.  In this, we again take our example from the
seL4 approach, which used just such an \emph{executable
specification}~\cite{Derrin:2006:RMA} to prototype the kernel prior to
implementation in C.

Given our target environments of Linux and Barrelfish, operations and data
structures for the reference monitor are named in a manner suggestive of an OS
kernel, although other implementations would be possible.  The most important
detail added at this stage is to make \textbf{translation structures}
explicitly visible.  The reason for this is to allow us to express the fact
that the translation state of the system depends, in a deterministic manner,
on the contents of RAM and device registers (e.g.~segment registers).  This in
turn allows us to express the invariant (necessary for integrity of the
reference monitor) that no such objects are ever made accessible (i.e.~mapped)
outside the monitor itself:
\invariant{obj:noaccess}{Never Accessible}{\\
Subjects can never access unmappable objects}

\begin{figure}
  \begin{center}
    \includegraphics[width=\columnwidth]{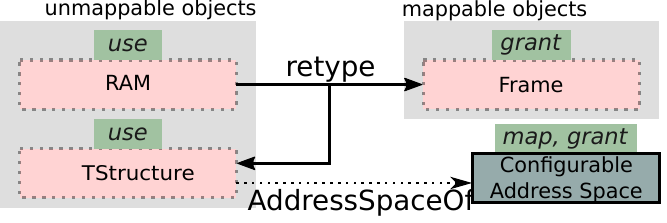}
    \caption{Object Type Hierarchy and possible rights (green).}
    \label{fig:objecttypes}
  \end{center}
\end{figure}

Note that (in contrast to the seL4 executable specification), the details of
the translation structures are kept opaque at this point---we merely record
that they exist at certain locations by dividing the mappable address spaces
into \emph{objects} (with terminology borrowed from Barrelfish):

\begin{footnotesize}
\begin{verbatim}
data Object = RAM {base :: Name, size :: Natural} 
            | Frame {base :: Name, size :: Natural} 
            | TStructure {base :: Name, size :: Natural} 
\end{verbatim}
\end{footnotesize}

Objects form a hierarchy (\autoref{fig:objecttypes}) which defines how objects
can be \emph{derived} from each other.  For example, translation structures
(\texttt{TStructure}) are created by retyping RAM objects.  The previous
invariant now reduces to stating that no object of type \texttt{TStructure} is
ever mapped.  \texttt{RAM} is the base type for untyped memory, and a
\texttt{Frame} is RAM that has been retyped to be mappable.

In addition, the set of translation structures defines (again in an
implementation-specific manner), the set of address spaces:

\begin{footnotesize}
\begin{verbatim}
AddressSpaceOf :: TStructure -> AddressSpace
\end{verbatim}
\end{footnotesize}

Authority is likewise stored as explicit rights:

\begin{footnotesize}
\begin{verbatim}
data Authority = Access Object | Map Object 
               | Grant Authority
\end{verbatim}
\end{footnotesize}

The monitor (kernel) state is a set of subjects (the term dispatcher being
borrowed from Barrelfish), a \emph{mapping database} (MDB) recording the
derivation relation between objects, and a set of active address spaces:

\begin{footnotesize}
\begin{verbatim}
data KernelState 
     = KernelState (Set Dispatcher) MDB (Set AddrSpace) 
\end{verbatim}
\end{footnotesize}

\begin{figure}
\begin{footnotesize}
\begin{verbatim}
mappingTrace :: (Operation KernelState)
mappingTrace = do
    ...
    -- retype a RAM object to a Frame    
    res <- retype RAM Disp Frame Disp
    -- retype another RAM object to a translation structure
    res <- retype RAM2 Disp TStructure Disp
    -- map the frame into the translation structure
    mapping1 <- Model.map TStructure Frame Disp
    ...
\end{verbatim}
\end{footnotesize}
\vspace{-0.5cm}
\caption{Mapping a RAM object}
\label{fig:model:trace}
\end{figure}

The monitor is exercised (as for the seL4 specification) by direct calls to
its API, such as in \autoref{fig:model:trace}.  These are implemented within a
state monad: \verb|Operation|. Thus changes to the system's state are a
sequence of API calls e.g.~retype or map:

\begin{footnotesize}
\begin{verbatim}
data Operation a = Operation (State -> (a, State))
  instance Monad (Operation) where ...
\end{verbatim}
\end{footnotesize}

Traces are thus sequences of such operations, corresponding to an observed
sequence of \verb|KernelState|s.  Each of these states defines a static
configuration of the decoding net.  Operations include:
\begin{itemize}
    \item \texttt{retype} converts an existing object into an object of a
    permissible subtype.
    \item \texttt{map} installs a mapping in a translation structure.
    \item \texttt{copy} copies the rights from one subject to another.
\end{itemize}

Contained within the set of all possible traces $T$, there is a set of correct
traces $CT \in T$ that correspond to sequences of consistent
\verb|KernelState|s. All other traces indicate that execution had to be
aborted at some point since an operation was applied that would otherwise have
led to transitioning to an inconsistent or disallowed system state.



\section{Implementation}
\label{sec:implementation}

We first describe how a subset of our model might be implemented in
the Linux monolithic kernel, and then present a full implementation 
based on the open-source Barrelfish OS~\cite{Baumann:2009:MNO}.   We
refer to this new implementation as \system, where MAS refers to Multiple 
Address Spaces.


\subsection{Implementation in a Monolithic Kernel}
\label{sec:impl:linux}
We describe how one could implement the least privilege model and add support 
for multiple address spaces in a monolithic kernel at the example of Linux.

The Linux kernel acts as the reference monitor and therefore assumes authority 
over all address spaces in the system: it can change address space mappings and 
grant access to memory at will. This happens mostly as a reaction to user-space 
requests such as \verb|mmap|, but may also originate from policy decisions inside 
the kernel e.g. demand paging or page caches where the kernel decides to unmap 
memory from a process.

A possible way to achieve separation is through intercepting updates to the 
translation tables, which can be done using the para-virtualization subsystem. 
Whenever a translation table is changed, this gets converted into an API call to 
the reference monitor. This gives some form of separation, but without proper 
virtualization cannot be strictly enforced.

User-space processes may share memory by creating shared memory objects, which 
are implemented as files in a ramfs. Linux manages access to that shared memory 
object--and file-based objects in general--using standard UNIX permissions, 
representing an access control list. Consequently, every process with a matching 
user or group id can access the shared memory object. ACL means for each object 
(resource) in the system there is a list of subjects plus rights. Files can be 
opened, which gives the process a file descriptor, which can be \verb|mmap|ed, 
hence a read right on a file can be seen as a grant right to the memory 
described by that file. After opening a file, the file descriptor can also be 
passed around, hence also the file descriptor represents a grant right. 

Most memory used by applications is not file backed, and hence referred to as 
anonymous memory. User-space processes have the access right to mapped anonymous 
memory. The process cannot explicitly hand over the grant right to anonymous 
memory to another process other than \verb|fork|ing itself, where the child 
process inherits rights on resources from its parent.

A process can request memory to be mapped and unmapped from its address 
space. It may supply hints on what type of memory it would like, but in 
the end the Linux kernel decides where to map and what memory to grant. 

%

Apart from tracking rights, our design also requires the understanding of
multiple address spaces and have rights refer to qualified names instead of addresses.
In order for Linux to do this, we have to make sure that
the physical frames are correctly identified in the presence of multiple address
spaces (e.g. the kernel sees a RAM region at a different address than say a DMA
engine). Each frame is identified by a physical frame number (PFN). We can use
this PFN as the canonical name for the frame itself and the sparse memory model
in Linux \cite{lwn:2005:linuxsparsemem} to implement multiple address spaces
holding physical resources as memory sections. For each frame of memory, Linux 
maintains a data structure tracking its use. We can augment the data structure 
to include type information to implement different memory object types (Linux 
already distinguishes between user and kernel objects). The relationship between 
PFNs and local physical addresses would need to be changed from a fixed offset 
to one that depends on the current configuration of translation hardware plus 
the current executing core.

In conclusion,  the Linux kernel acts as a central authority holding the grant 
and map right to all address spaces and resources. A separation is possible when
using the para-virtualization subsystem to intercept updates to translation 
tables. The kernel data structures could be modified to support the
notion of multiple address spaces. Because it is very hard to support the full
granularity of our model in a monolithic kernel, we chose to implement and
evaluate it in a capability system, which is described in the next section.




\subsection{Implementation in \system}
\label{sec:impl:barrelfish}

We chose the open-source Barrelfish OS~\cite{Baumann:2009:MNO} as the basis 
for our implementation because it uses an seL4-style capability system
for authorization and resource management, but in contrast to seL4 has
has support for heterogeneous platforms and has drivers for IOMMUs 
and the Intel Xeon Phi co-processor, thus providing a real-world
example of complex addressing. 

We describe the relevant parts of our implementation in \system: the
capability system that supports multiple address spaces
(\autoref{sec:impl:capabilities}), implementation of
runtime support by generating code for known translation and maintaining a graph
of configurable nodes \autoref{sec:impl:runtime}, and finally adapting
user-space device drivers (\autoref{sec:impl:devicedriver}). 

\subsubsection{Capability System}
\label{sec:impl:capabilities} 

Barrelfish manages physical resources using a capability system for
naming, access control, and accounting of objects in a single physical
address space.  We describe the \system capability system as a whole
here, since a clear description of the original Barrelfish capability
system has not been published.   As in seL4~\cite{Elkaduwe:2008:VPM},
capabilties are \emph{typed} to indicate what can be done with the
memory they refer to; rules dictate valid \emph{retype} operations
(e.g retyping RAM to a Frame).  

\system builds on Barrelfish by adding multiple address
spaces and having capabilities which refer to memory objects
hold the object's canonical base name, the size of the object  
they are referring to, as well as its type and rights. 

\system is a \emph{partitiioned capability system}: Capabilities are
stored in memory-resident objects as well, but these are
\textit{unmappable} ensuring that no user-space process can forge
capabilities by writing to memory locations.  A process holding a
capability obtains a certain set of rights on the object referred to
by the capability. These rights can be exercised by invoking the
reference monitor API which is implemented as a system call interface.


Capabilities encode the canonical names of the objects they refer to,
implemented as a struct with two fields: the address space identifier
(ASID) and the address within the address space. An optimized variant  
packs both values into a 64-bit integer providing support for a 16-bit ASID and 
a 48-bit address, which is sufficient for current platforms.

ASIDs nevertheless are a limited resource, and their allocation must be managed 
accordingly to avoid ASID exhaustion. We use a dedicated capability to manage 
ASIDs, where a new range of ASIDs can be allocated by retyping a larger range of 
ASIDs. 

There may be multiple capabilities pointing to the same object, but there is 
always at least one capability for every given byte in memory.  

\paragraph{The mapping database:}

\system manages a \emph{mapping database}, a data structure that allows 
efficient lookup of all related capabilities given the name of object they refer
to.  The mapping database is a balanced tree structure of all
capabilities present in the database.  

The mapping database stores the capabilities in a cannonical ordering,
allowing efficient lookup and range query operators such as ``overlap'' and 
``contains''. The canonical ordering of the capabilities is defined on their 
canonical name (address space and address), size and
type. Capabilities to objects with a smaller name  
appear first. If the base names of two capabilities are equal, then the larger 
object comes first.  Finally, all other attributes being equal, the
type of the capability defines the order: types higher up in the
hierarchy come first.  

This ordering is important, because based on the canonical order of the 
capabilities one can define the \emph{descendant} relation. We say a capability 
$B$ is a descendant of capability $A$ if $A$ is smaller than $A$ and $B$ is 
fully contained in the range convered by $A$:
\begin{align*}
\texttt{descendant}\ c_1\ c_2 \leftrightarrow c_1 \cap c_2 = c_2 \land c1.type 
\leq c_2.type
\end{align*}
The mapping database can therefore be traversed to find the descendants of a 
capability (successors) and ancestors (predecessors) efficiently.

It is important that the ordering relation is in line with the retype operation.
If $B$ can be retyped from $A$, $B$ must be smaller than $A$. Our
definition fulfils this, a retype can increase the name, decrease the size
or change the type to a subtype. 


With help of the mapping database, we can efficiently find all the ancestors 
and descendants of a particular object.

\paragraph{Page tables and address spaces:}
\system has a distinct capability type for each hardware-defined translation 
table e.g.\ one for each of the four levels of the x86\_64 architecture.
Each of these capability types are translation structures in the sense of the 
executable spec. 

User processes can construct their own page tables through capability 
invocations. This is safe, because the invocations only allow operations 
resulting in correct-by-construction page tables, and processes can only map 
resources for which they hold a capability with the grant right to it.

Since a page table defines an address space, we can \emph{derive} an 
\textit{address space} capability from a page table. This address space 
represents the input address space of the translation table. For each 
translation table, the spanning address space can only be derived once.

When we delete a page table, we use this stored ASID to query the
mapping database for address space capabilities and start a recursive deletion. This ensures that 
upon deletion of the page table, the address space is deleted
including all \textit{segments} within it.  This is equivalent to
\emph{revoking} all descendants of the address space capability and
then deleting it.

\paragraph{Tracking mappings:}
When access to an object is revoked, all positions where this object has been 
mapped must be found and removed.  We manage this bookkeeping using
the capability system.  
For each mapable object there exists a corresponding \textit{mapping} 
capability. The mapping capability is a descendant of (retyped from)
the mapped objects and hence we can find all locations where an
object is mapped by walking the mapping database in ascending
order. Each mapping capablity indicates the  
page table objects and slot range where the object has been mapped.

The same technique is used to track mappings of multi-level page
tables. For each valid entry in a page table there exists a mapping
capability. When the last mapping capability is deleted, the
page table entry is invalidated.



\subsubsection{Runtime Support}
\label{sec:impl:runtime} 
In~\autoref{fig:model:multiple-address-spaces} we draw a diagram of the 
different address spaces present in a heterogeneous multi-processor system. To 
acquire the access right to a particular memory object, a sequence of 
translations need to be setup. Which address spaces need to be 
configured depends on the system topology, which may only be discovered at 
runtime. 

\paragraph{SoC-Platforms} 
The topology of SoC platforms is typically fixed and known at compile time. 
We can therefore enumerate all address spaces of the SoC and pre-compute all 
fixed translations and store a graph of the topology consisting of configurable 
and leaf address spaces in the kernel. We can \emph{generate} core-specific 
translation functions that convert local addresses to global 
names and vice versa. The name can then be resolved by walking the translation 
structures of the configurable address spaces until it reaches an accepting 
address space or there is no translation. We evaluate this scenario 
in~\autoref{sec:eval:simulators}.

\paragraph{Device Discovery}
In general, the information about the hardware topology and its address spaces 
may be incomplete and must be discovered during runtime. For instance, the 
presence of an IOMMU is known after parsing the ACPI tables and the Xeon Phi 
co-processor of our example (\autoref{fig:model:multiple-address-spaces}) is 
discovered by PCI, and lastly the size of the GDDR available on the co-processor 
is known by the driver. The state of the model is therefore populated by 
multiple sources of information.

In Barrelfish, there exists the system knowledge base 
(SKB)~\cite{Schupbach:2011:DLA} which stores information about the system. The 
SKB in a nutshell is a database storing facts about the system which can be 
queried using Prolog. We implement the model inside the SKB. During device 
discovery, processes insert information about the discovered address spaces and 
how they are connected with each other.

\paragraph{Model Queries}
Device drivers must configure translation units to enable devices to access 
memory. Booting a core on the Xeon Phi co-processor is a particular example: 
application modules to be run on the co-processor may reside in host RAM. To 
make this accessible from the co-processor the IOMMU and the SMPT must be 
configured accordingly. This information can be obtained by querying the SKB, 
which returns a list of address spaces that must be configured. The query is 
based on a shortest path algorithm between the address space of the Xeon Phi 
core and the address space where host RAM resides in. 

Running the queries in the SKB is costly (\autoref{sec:eval:scaling}). We provide a 
library that caches the graph representation of configurable address spaces 
and run shortest path on it. 

The result of the query is a list of address spaces that need to be configured to 
make the memory object accessible from the source address space. This blueprint 
is then converted by the user-space process into a sequence of capability 
operations to allocate memory, setup translation structures and perform the 
relevant mappings. The model queries only provide a `hint' on what needs to be 
configured while the capablity system enforces the authorization required to 
perform the required mappings. We evaluate the latency of this scenario in 
~\autoref{sec:eval:realhw}.

\paragraph{Address Resolution}
While the SKB stores the address space topology of the system it does not store 
the actual translations of configurable address spaces. An address can be fully 
resolved by performing the previous query and instead of changing the 
configuration of the address spaces, we can use the translation structure to 
calculate where the address space translates the address.

\subsubsection{Device Driver Adaptation}
\label{sec:impl:devicedriver} 

We adapt the user-space device drivers in Barrelfish to use the runtime support
described above when configuring their devices and allocating in-memory data
structures.  In \system, device drivers run in user-space. They are started by
a device manager which passes a set of capabilities including a capability to
the device registers and the IOMMU IPC endpoint. The driver can then use 
capability operations to map the device registers into its address space or 
program the IOMMU translation through the IOMMU IPC endpoint.  Devices with 
additional memory, such as the Xeon Phi with GDDR receive a capability to the 
leaf address space, which the driver can then use to retype new RAM capabilities 
from it. 

Memory access from the device might be translated by the IOMMU. To setup a
shared buffer between the driver and the device the driver needs to: Allocate
memory, Map the memory into the driver's own address space, query the graph to
determine necessary configuration steps, follow the result to map the memory
into the device's address space For an evaluation of these steps
see~\autoref{sec:eval:realhw}. 

To set up the IOMMU, we implemented two alternatives: \ei an RPC to the 
IOMMU reference monitor that manages the translations, or \eii direct capability 
invocations on the translation table used by the IOMMU for this device. This is 
safe, because the capability system enforces that only memory for which the
driver has a capability for can be mapped.

\section{Evaluation}
\label{sec:evaluation}

We evaluate our implementation by showing memory 
management performance comparable to Linux (\autoref{sec:eval:memops}) and 
applicability to a real-world scenario using co-processors 
(\autoref{sec:eval:realhw}).  We also show the scaling behavior of the model queries 
(\autoref{sec:eval:scaling}) and demonstrate how the model can also be used in 
pathological topologies using simulators (\autoref{sec:eval:simulators}).
Finally, we analyze the space-time overheads of our implementation 
(\autoref{sec:eval:overheads}).

All performance evaluations use a dual-socket Intel Xeon E5 v2 2600 
(``Ivy Bridge'') with 256GB of main memory.  There are 10 cores per socket with 
HyperThreading, TurboBoost, and speed stepping disabled, and the
system runs in ``performance'' mode.  The system also has two Intel Xeon Phi co-processors 
(``Knights Corner'').  All Linux experiments use Ubuntu 
18.04LTS, with kernel version 4.15 and the latest patches for
mitigating Meltdown and Spectre attacks.


\subsection{Memory operations}
\label{sec:eval:memops}

We compare the performance of \system's memory subsystem against Linux
with Spectre/Meltdown mitigation both enabled and disabled, using two
microbenchmarks.  \system has no mitigation measures.

\subsubsection{The Appel and Li benchmark}
\label{sec:eval:memops:appel}
~\cite{Appel:1991:VMP} tests operations relevant to 
garbage collection and other non-paging tasks by measuring time to 
protect, and trap-and-unprotect pages of memory.

We run the benchmark with working sets of less than 2MB (512 pages).
We measure Linux with four configurations: \ei default TLB flush
heuristic, and \eiii always full TLB flush, all with Spectre/Meltdown
mitigation both enabled and disabled. We benchmark \system in two
ways: \ei direct invocation of the mapping capability and \eii
protecting the page through user-level data structures tracking the
mapping. Note that \system does not support selective TLB flushing.

\begin{figure}[t]
    \begin{center}
        \includegraphics[page=1,width=\columnwidth]{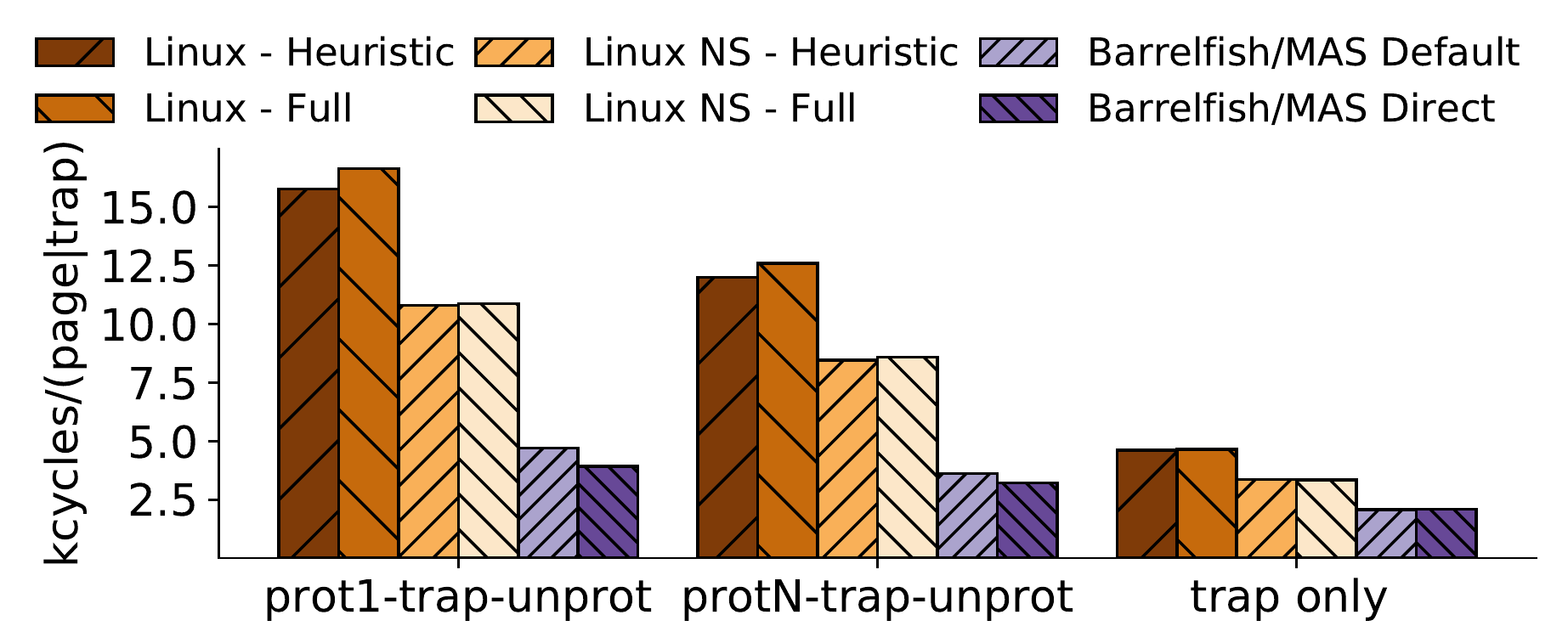}
    \end{center}
    \vspace{-0.5cm}
    \caption{Appel-Li benchmark on \system and Linux with and without
    Spectre/Meltdown mitigation (NS). }
    \label{fig:appel-li}
\end{figure}

%

The results are shown in~\autoref{fig:appel-li}. We observe that 
\system is consistently faster than Linux in all cases. The Spectre/Meltdown 
mitigation incurs a 45-53\% slowdown. For both multi-page 
(\emph{protN-trap-unprot}) and single page  (\emph{prot1-trap-unprot}) 
protect-trap-unprotect, \system is up to 4x faster than Linux. We observe a 
slight increase in execution time when full TLB flushes 
are enabled. The \system ``Direct'' results use the kernel primitives directly.
This enables us to isolate the cost of user-space accounting, which accounts for 
10-17\% of the execution time.

\subsubsection{The map/protect/unmap benchmark}
\label{sec:eval:memops:vmops}
measures the performance of the primitive operations \verb|map|, 
\verb|protect| and \verb|unmap| with respect to an increasing buffer size.

The benchmark works as follows: \ei allocate a region of virtual 
memory and fault on it to \verb|map| memory, \eii write-protect the entire 
virtual region, and \eiii unmap the virtual memory region again.  We time each 
operation separately. We measured different ways to map memory on Linux using 
\verb|mmap|, \verb|shmat| and \verb|shmfd| and compare \system against the 
\emph{best} performance we obtained on Linux for each operation and page-size.
For mapping and unmapping 4kB pages, this was passing a file descriptor obtained through
\verb|shm_open| to \verb|mmap|. For map/unmap with larger page sizes,
shared memory segments (\verb|shmat|, \verb|shmdt|) performed best.
Changing page protection was always fastest using \verb|mprotect|.
Again, we benchmark Linux with and without Spectre/Meltdown mitigation enabled. 
If possible, we do not measure the time for memory allocation as this is 
dominated by \verb|memset|. On \system we use the high-level interfaces to 
include user-space book-keeping in the measurements.

\begin{figure*}[ht]
  \begin{center}
    \includegraphics[width=\textwidth]{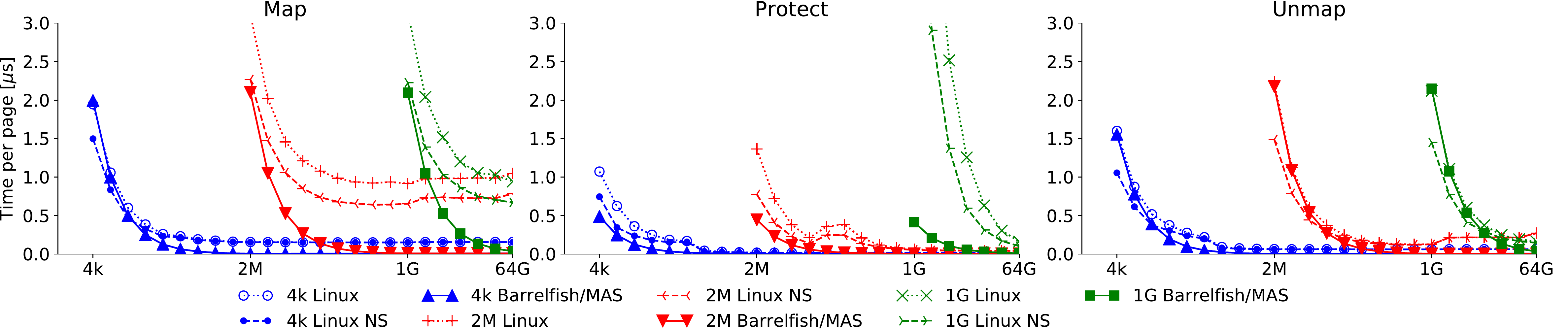}
  \end{center}
  \vspace{-0.5cm}
  \caption{Comparison of memory operations on \system and Linux with and without 
  Spectre/Meltdown mitigation (NS). Execution time per page in $\mu$s. Buffer 
  sizes in powers of two from 4kB to 64GB.}
  \label{fig:results:memops-compare}
\end{figure*}

\autoref{fig:results:memops-compare} shows execution time of the three 
operations per page for an increasing buffer size and three 
page sizes. Enabling Spectre/Meltdown mitigation results in a
slow down of up to 2x for small page numbers. In all cases, the cost per page 
decreases as the number of pages increases, amortizing the system call cost.

\textit{Map:} \system is able to match and outperform Linux in all but one 
case, with a significant difference when using large and huge pages. 

\textit{Protect:} These are in line with the Appel and Li benchmarks
above; \system outperforms Linux in all configurations.

\textit{Unmap:} We observe very similar performance characteristics here.
For small buffer sizes Linux is slightly faster, for larger buffers \system
slightly outperforms Linux.


From these two microbenchmarks, we conclude that \system memory
operations are competitive: capabilities and fast traps allow an
efficient virtual memory interface despite splitting up larger
mappings into multiple capability operations and syscalls.  
It is possible to build a fast and competitive memory system which
still fully implements our fine-grained, least-privilege model.


\subsection{Complex hardware}
\label{sec:eval:realhw}

We now profile the support for address space networks in \system,
including memory mappings and model queries. We put the cost in the
context of related operations a device driver has to perform.

We profile the boot process of the Xeon Phi Co-Processor on our server platform. 
All accesses from the co-processor to host RAM are translated multiple
times, most notably:
$$Core MMU \rightarrow SMPT \rightarrow IOMMU \rightarrow System Bus$$
Each step must be configured correctly. We adapted the existing drivers for the
co-processor, the system memory page table (SMPT), and the IOMMU to use our new
capabilities and model queries. The MMU is managed by the kernel
running on the co-processor cores. Resources are managed using the capability 
system which allows safe programming of translation tables. 

First, we allocate 6MB from host RAM, and map this into the device
drivers address space (equivalent to performing an anonymous
\texttt{mmap} in Linux).  Then we copy the boot image into this
allocated buffer.  We then query the model representation to determine which
translation units must be reprogrammed.  We map the buffer into the
device's IOMMU address space, and then map the resulting 
obtained segment into the SMPT address space.  We compare the IOMMU
mapping in two cases. In the first, we ask the IOMMU driver to
perform the mapping; this ad-hoc approach corresponds to the current
state of the art.  In the second, enabled by our model, we perform the
mapping directly with capability invocations.

\begin{figure}
  \includegraphics[width=\columnwidth]{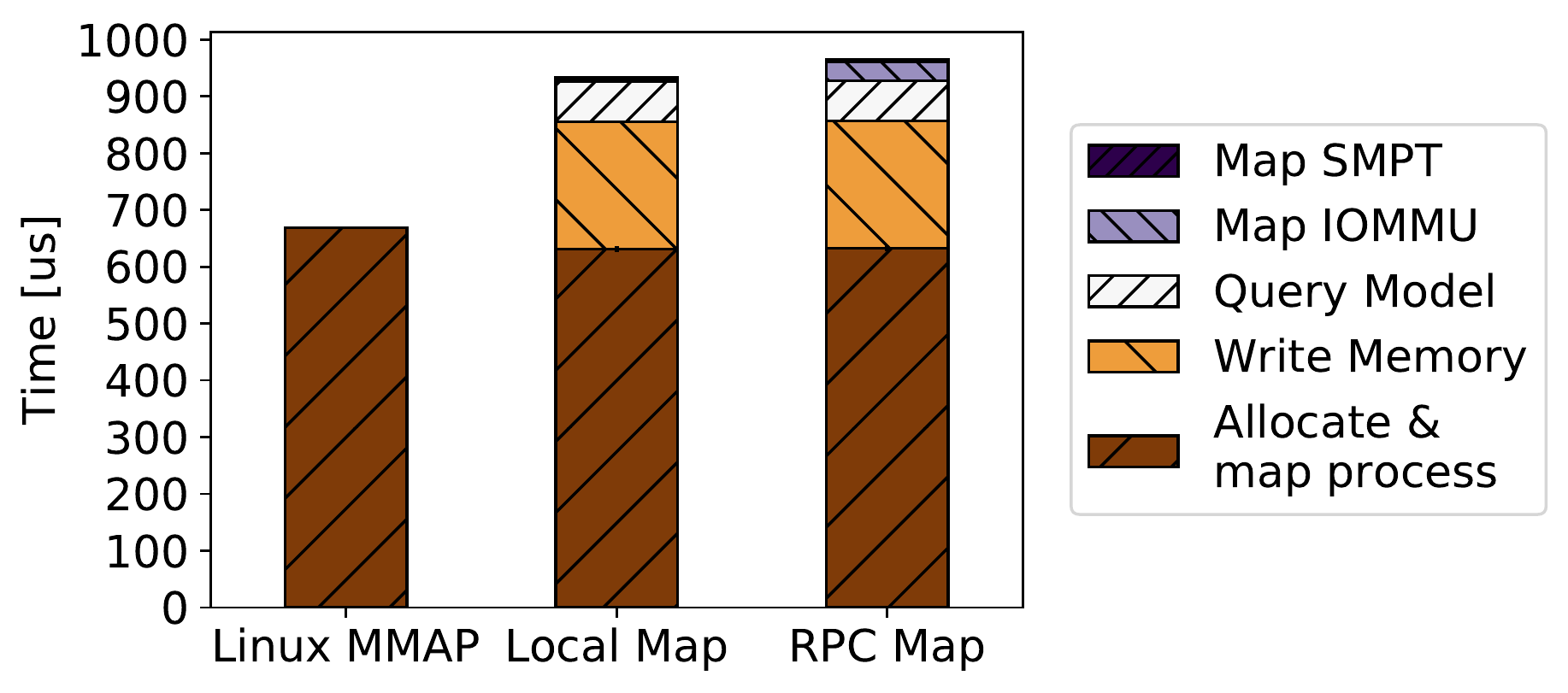}
    \vspace{-0.5cm}
  \caption{Profiling Configuration Time for a Xeon Phi Co-Processor comparing 
  Local Syscalls and RPCs to Perform IOMMU Mappings with Linux 
  \texttt{mmap}'ing a buffer of the same size for perspective.}
  \label{fig:eval:bufalloc}
\end{figure}

As \autoref{fig:eval:bufalloc} shows, the cost is dominated by memory
allocation, which takes about 625$\mu$s and involves an RPC to a memory server. 
Writing the buffer content using \texttt{memcpy} takes 224$\mu$s.
Determining the units to be configured to make the buffer
available to the device takes 71$\mu$s, using the C graph implementation. \LH{
    make sure the implementation explains the graph.}
Setting up the IOMMU mapping is 2$\mu$s (or 32$\mu$s, when using RPC). Mapping the
segment into the SMPT using a kernel driver takes 5$\mu$s. 


For comparison we also show the cost in Linux to \texttt{mmap} an anonymous
6MB region in a userspace process -- equivalent in our
implementation to allocating and mapping the buffer in the driver. We perform
this operation slightly faster than Linux, but pay an additional
71$\mu$ to dynamically determine the nodes that have to be
configured.  A less flexible approach might pre-compute or memoize
this step, avoiding the latency at map time.  Compared with
the cost of allocation and writing the memory, the cost of setting up the IOMMU
and SMPT mappings (together 7$\mu$s) are negligible. Note that \textit{in any
system} an untrusted agent will have to perform some sort of
invocation (such as a system call) to install these mappings. Despite
our fine-grained rights and dynamic implementation, performance is
comparable to Linux.


\subsection{Scaling}
\label{sec:eval:scaling}

We now turn to the scaling properties of the model
representation with respect to the system complexity.  In real
systems, we see ever-increasing numbers of cores and DMA-capable
devices, but the diameter of the decoding net representation grows
much more slowly, and rarely exceeds 10.  This is true not only for
x86 systems, but also for all the ARM SoCs we have encountered to
date. 

We write a synthetic benchmark that simulates a system with an increasing
number of PCIe devices, each of which has its own address space and
translation unit, much like the Intel Xeon Phi described in the
previous section.  This grows the model state in two ways: the total
number of address spaces, as well as the number of these that are 
configurable.  Both grow linearly with the number of PCIe devices. We
measure the time it takes to determine the configurable address spaces
between a PCIe device and the system bus, a typical setup operation
from a device driver that has to setup IOMMU and 
device-local translation structures.  We evaluate two implementations: \ei a
Prolog implementation of the model using the EclipseCLP interpreter in
Barrelfish, and our C implementation based on a graph represented
as an adjacency matrix.  Both use Dijkstra's algorithm on the graph
representation.  

\begin{figure}
  \includegraphics[width=\columnwidth]{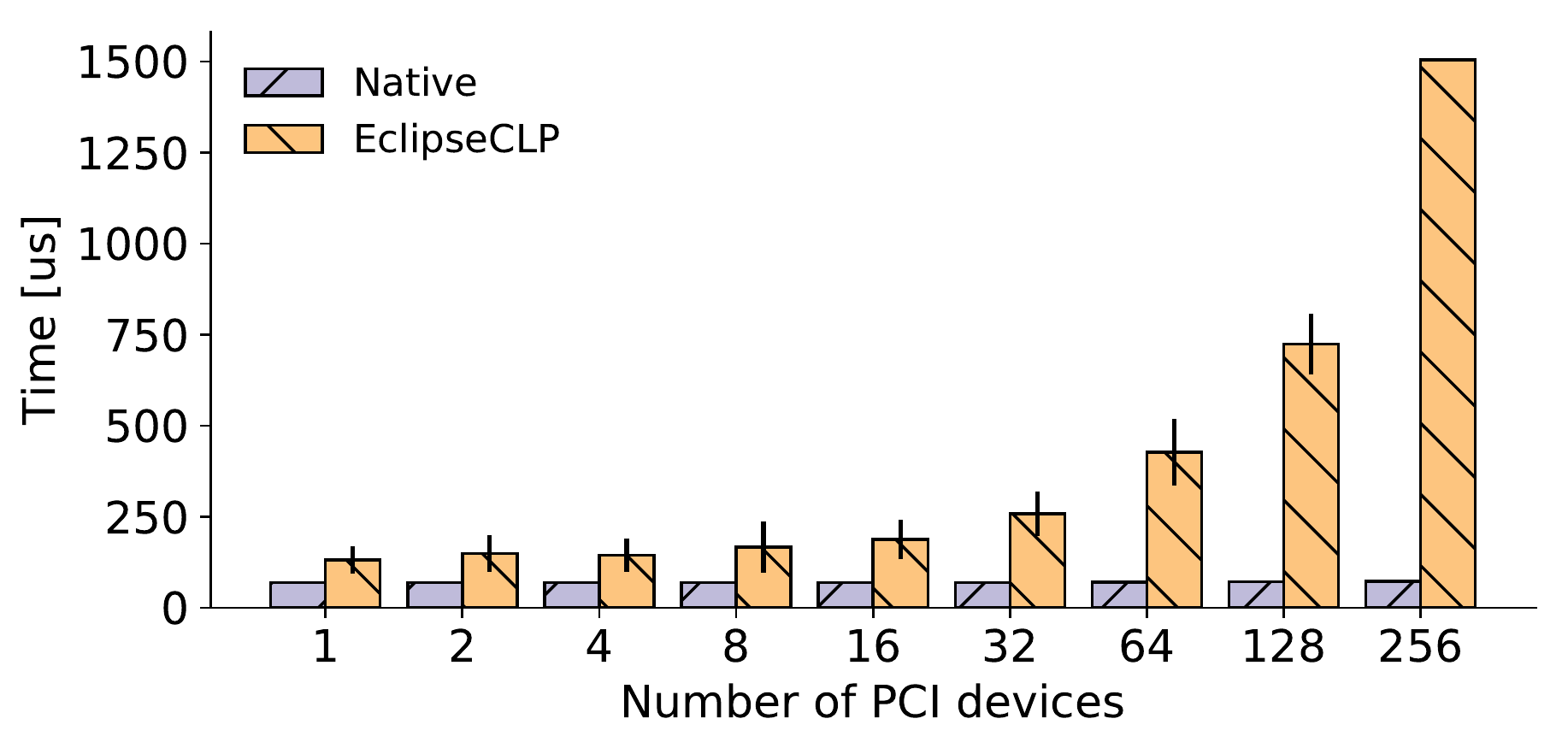}
  \caption{Cost of determing mappable nodes on an X86 system with growing
    number of PCI devices. \RA{update plot}}
  \label{fig:eval:x86}
\end{figure}

\autoref{fig:eval:x86} shows that, due to internal memory allocations,
the performance of EclipseCLP implementation scales linearly in the
number of devices. The native C implementation, in contrast, shows
almost constant performance.  Small linear factors stem from walking
of the adjacency-matrix and initialization of the parent array.

We conclude that the cost of determining configurable nodes remains almost independent of the
system complexity, as long as the graph is of low diameter and is
maintained in an efficient data structure, suggesting that the routing
calculation is feasible for modern hardware.


\subsection{Correctness on simulated platforms}
\label{sec:eval:simulators}

In this qualitative evaluation we show that the model implementation
is functional and performant even when run on simulated platforms with
unusual address space topologies not supported by other systems.
While these topologies are extreme, their envelope includes other real
systems (such as those with secure co-processors) which are not
handled by current systems. 

We wrote a series of system descriptions for the ARM Fast Models
simulator~\cite{fastmodels}. We use this  
description to \ei configure the simulator and \eii extract the topology of the 
memory subsystem. We then use this information to populate the address space 
model which is used at compile time to generate operating system code and 
at runtime to query information about the memory system as in the previous 
evaluation. We mention four configurations, where each consists of two 
ARM Cortex-A57 clusters, each having their own memory map connecting to DRAM and 
other devices. The memory map is configured as follows:
\begin{enumerate}
    \item \emph{Uniform} Uniform memory map between all clusters.
    \item \emph{Swapped} Memory map contains two areas whose addresses
      are swapped (exchanged) between the two clusters.
    \item \emph{Private} Each cluster has its own private memory region.
    \item \emph{Private Swapped} A combination of the Private and Swapped 
    configurations
\end{enumerate}

We know of no other current OS designs which can manage memory
globally in all these cases.  Popcorn Linux~\cite{Barbalace:2015:PBP}
and Barrelfish have limited support for case 3; while regular Linux
and seL4 only support case 1. 

\system is able to boot and manage memory on all platforms without
modifications, regardless of the topology, by virtue of the
capabilities used to refer to memory containing the canonical name of
the object.  Whenever an object is accessed, this canonical name is
converted into a local address using a generated function.


\subsection{Space and time overheads}
\label{sec:eval:overheads}

Finally, we analyze the time and space complexity of managing the 
physical resources of the system using capabilities in Barrelfish.

We are interested in the space overhead to store the capabilities, managing the 
lookup of capabilities in the mapping database, and creating new mappings. 

In the implementation, capabilities occupy 64 bytes each. There is
typically less than one capability for each frame of memory, as each
capability can represent up to $2^{64}-1$ bytes of memory.  The
number of capabilities should grow sub-linearly with the size of
available RAM.  
For each mapping, \system creates a capability for bookkeeping. The number of these
mapping capabilities also grows sub-linearly in the total number of mapped 
frames.  Large frames result in one mapping capability per page-table that 
is spanned for the mapping.
Since the 64-byte capability representation also includes all the
pointers necessarily to index the mapping database, the latter incurs
no additional overhead.  The index itself is a balanced tree; lookups
are logarithmic in the total number of capabilities. 


Overall, keeping track of memory resources with capabilities incurs 
a space overhead which grows at worst linearly in the available
physical memory.  Furthermore, the mapping database can be implemented and queried 
efficiently. Assuming the \system worst case of one capability per 4kB frame,
this accounts for a 1.5\% total memory overhead. In comparison, Linux manages a
\texttt{struct page} per physical frame of up to 80 bytes in size, an
overhead of almost 2\%.


\section{Conclusion}
\label{sec:conclusion}

In this paper we have built on existing work in modelling the complex
interacting address spaces in modern hardware by adopting the proven
methodology of the seL4 project to produce a rigorous, no-stone-left-unturned
model of memory management.  Our model applies well-known concepts in access
control, giving an abstract model amenable to implementation in
capability-based systems (e.g.~Barrelfish), as well as ACL-based systems such
as Linux.

We have shown that it is possible to implement the model efficiently in an
operating system delivering excellent memory management performance while at
the same time offering a clean and safe way to deal with the complexity of the
allocation and enforcement problem.

We've shown that the model can be used to configure real, complex (even
pathological) systems, scales well, and introduces little overhead.  Our
model is a sound foundation for both fully verified systems and more reliable
memory management in existing systems.

%
%

\bibliographystyle{plain}
\bibliography{references}

\end{document}